\documentclass[twocolumn]{jetpl}
\usepackage{graphicx}
\usepackage{amssymb}
\usepackage{amsmath}
\usepackage{color}
\usepackage{multirow}
\usepackage{hyperref}

\title{The effect of pressure on the splitting in $^3$He in nematic aerogel}
\rtitle{The effect of pressure on the splitting in $^3$He in nematic aerogel}
\sodtitle{The effect of pressure on the splitting in $^3$He in nematic aerogel}

\author{V.\,V.\,Dmitriev$^+$, D.\,V.\,Petrova$^{+\circ}$, A.\,A.\,Soldatov$^+$\thanks{e-mail: soldatov\underline{ }a@kapitza.ras.ru}, A.\,N.\,Yudin$^+$}
\rauthor{V.\,V.\,Dmitriev, D.\,V.\,Petrova, A.\,A.\,Soldatov, A.\,N.\,Yudin}
\sodauthor{V.\,V.\,Dmitriev, D.\,V.\,Petrova, A.\,A.\,Soldatov, A.\,N.\,Yudin}

\address{$^+$P.\,L.~Kapitza Institute for Physical Problems of RAS, 119334 Moscow, Russia}
\address{$^\circ$HSE University, 101000 Moscow, Russia}

\dates{\today}{*}

\abstract{Here, we present the results of an investigation of how the pressure affects the splitting of the superfluid transition temperature in $^3$He in anisotropic aerogel. It is well known that boundary conditions significantly influence the properties of superfluid phases in aerogel. When aerogel strands are coated with a $^4$He layer in magnetic field, new phases, such as the polar, polar-distorted A (DA), polar-distorted B (DB), and $\beta$ phases, become energetically favorable. In contrast, without this coating, the system tends to favor phases resembling the bulk A, B, and A$_1$ phases. Our earlier results showed a nonlinear dependence of the superfluid transition temperature splitting in pure $^3$He, but the range of nonlinearity did not match the theoretical predictions based on magnetic scattering effects. To further investigate this discrepancy, we performed measurements of the splitting under varying pressures for both pure $^3$He and with $^4$He coverage of the aerogel strands.}

\begin{document}

\maketitle

\section{Introduction}

The rich phenomenology of superfluid phases in $^3$He stems from its triplet Cooper pairing mechanism, which supports multiple quantum states with distinct symmetry properties \cite{VW}. In bulk systems, two primary phases emerge at zero magnetic field: the anisotropic A phase with nodal gap structure and the fully gapped isotropic B phase. The introduction of a magnetic field unveils a third phase (A$_1$), splitting the superfluid transition at $T=T_c$ into two stages: first, to the spin-polarized A$_1$ phase at $T=T_{A1}>T_c$, then to the A$_2$ phase at $T=T_{A2}<T_c$. This splitting exhibits linear field dependence ($1.2\text{--}6$\,$\mu$K/kOe) modulated by pressure \cite{Amb,isr84,sag84}.

Recent advances have shifted the focus to geometrically confined $^3$He, especially in aerogel matrices -- nanoporous networks where the anisotropy of the strands can significantly influence the phase realization. Weakly anisotropic aerogels preserve bulk-like A and B phases \cite{Osh,H01,dmit02,dmit10}, while strongly anisotropic ``nematic'' aerogels with globally aligned strands \cite{asad15} stabilize exotic states through directional quasiparticle scattering: the polar phase (non-chiral equal-spin paring state with a Dirac nodal line), distorted A and B phases (DA and DB) with hybrid gap structures, and the magnetic-field-induced $\beta$ phase (spin-polarized analog of A$_1$ phase) \cite{AI,fom18,fom20,sur19_1,dmit12,dmit14,dmit15,dmit18,dmit20,bet}.

Critical to this phase diversity are surface scattering effects at the aerogel strands. Specular reflection of $^3$He quasiparticles on $^4$He-coated strands preserves orbital coherence, enabling the polar and $\beta$ phases \cite{dmit15,bet}. On the contrary, in the absence of $^4$He (using pure $^3$He), the strands are covered with solid paramagnetic $^3$He layer making the scattering of quasiparticles diffuse, and phases similar to bulk ones are realized \cite{dmit18,dmit23,dmit25}.

Our previous experiments with pure $^3$He in nematic aerogel \cite{dmit23,dmit25} revealed anomalous deviation in the splitting of the superfluid transition temperature in magnetic field from a theoretical prediction regarding the range of fields where nonlinearity in the field dependence of the splitting should occur. This discrepancy motivates systematic investigation of pressure and boundary condition effects on the splitting, with and without $^4$He coverage.

\section{Theoretical models}
\subsection{Specular boundary ($^4$He coating)}
When cooling from the normal phase of $^3$He confined by a nematic aerogel, the superfluid transition to the $\beta$ phase should occur at a temperature:
\begin{equation}
T_{P1}=T_{ca}+T_c\eta H,
\end{equation}
where $H$ is the magnetic field, $T_{ca}$ is the superfluid transition temperature of $^3$He in aerogel at $H=0$, and $\eta\sim10^{-3}$\,kOe$^{-1}$ is the splitting coefficient \cite{sur19_1}.

With further cooling, a transition to a distorted $\beta$ phase is expected at a temperature:
\begin{equation}
T_{P2}=T_{ca}-T_c\eta H\frac{\beta_{12345}}{-\beta_{15}},
\end{equation}
where $\beta_{15}=\beta_1+\beta_5$, and so on, $\beta_i$, $i\in\{1,\ldots,5\}$ are coefficients in the Ginzburg-Landau free energy functional, or beta parameters \cite{VW}. Upon cooling, the distorted $\beta$ phase continuously transforms into a pure polar phase.

The splitting ratio is determined by Ginzburg-Landau coefficients:
\begin{equation}\label{beta}
\frac{T_{P1}-T_{ca}}{T_{ca}-T_{P2}}=\frac{-\beta_{15}}{\beta_{12345}}.
\end{equation}

\subsection{Spin-polarized impurity model (pure $^3$He)}
According to \cite{sh,bh}, the solid $^3$He coatings on the aerogel strands act as paramagnetic impurities, suppressing the splitting of the superfluid transition temperature. The upper ($T_{ca1}$) and lower ($T_{ca2}$) critical temperatures are given by:

\begin{equation}\label{tca12}
T_{ca1,2}=T_{ca}\pm\left[\eta_{1,2}-C_{1,2}\frac{\tanh(\alpha H)}{\alpha H}\right]H,
\end{equation}
where $\eta_{1,2}=\eta^0_{1,2}\cdot T_{ca}/T_c\sim1\,\mu$K/kOe \cite{isr84,sag84} are the splitting parameters in the absence of spin exchange ($\eta^0_{1,2}$ are those for the bulk A$_1$ phase), $C_{1,2}\sim\xi_0/l_s\sim1\,\mu$K/kOe is the spin-exchange parameter with $\xi_0$ the superfluid coherence length and $l_s$ the magnetic mean free path which represents the average distance between spin-flip scattering events. The field scaling factor $\alpha=\gamma\hbar/(2k_BT_{ca})=\alpha_0\cdot T_c/T_{ca}$ contains the gyromagnetic ratio $\gamma$ of $^3$He nuclei and Boltzmann constant $k_B$. Here, $\alpha_0$ corresponds to the case of zero suppression of a superfluid transition temperature.

According to Eq.~\eqref{tca12}, the nonlinearity region of the field dependencies of $T_{ca1,2}$ is controlled by the parameter $\alpha$ and should be essential at the fields $\le20$\,kOe, while at higher fields the splitting recovers linearity as the solid $^3$He magnetization saturates. In our recent studies with nematic aerogel, the nonlinearity region was about 6 times smaller than expected \cite{dmit25}. Yet, experiments with pure $^3$He in isotropic silica aerogel do not contradict the theory, although the nonlinearity was not observed because the experiments were carried out in magnetic fields either less than 8\,kOe (where the splitting was not detected) \cite{halp02} or greater than 70\,kOe (where the splitting was linearly dependent on $H$) \cite{halp04}.

\subsection{Spatial correlation effects (pure $^3$He)}
Unlike the mean-field theories \cite{sh,bh}, Surovtsev’s theory \cite{sur25} takes into account spatial correlations of the impurity magnetization. That is, it considers the effect of deviations from complete disorder of the medium associated with the spin degree of freedom. In zero magnetic field, maximal disorder leads to the maximum possible suppression of the transition temperature for a given concentration of magnetic impurities. At the same time, according to Surovtsev’s approach, correlated fluctuations of the order parameter induced by the aerogel increase the superfluid transition temperature to the A$_1$ phase compared to the case of a completely disordered medium.

The magnitude of the temperature increase depends on the correlation radius of impurity magnetization in the paramagnetic $^3$He film $R_s\sim100$\,nm, as well as on the magnetic mean free path in the aerogel $l_s\sim1000$\,nm (which accounts for spin-flip processes):
\begin{equation}
\delta\tau^{(1)}\left|_{H=0}\right.\sim\frac{R_s^2}{l_s^2}.
\end{equation}

Applying a magnetic field suppresses fluctuations of the spin part of the order parameter, similarly to how thermal fluctuations are suppressed when moving away from $T_c$ \cite{fom08}. Therefore, in a magnetic field, the correction $\delta\tau^{(1)}$ decreases. This effect is observed on top of the usual linear increase of the transition temperature to the A$_1$ phase in a magnetic field. Thus, in sufficiently strong magnetic fields, when the contribution of correlated fluctuations is completely suppressed, the dependence becomes linear again.

The magnitude of the magnetic field at which interpolation between the two regimes occurs is given by:
\begin{equation}
\eta H\sim\frac{\xi_0^2}{R_0^2}\cdot\frac{R_s^2}{l_s^2},
\end{equation}
where $R_0\sim100$ nm is the aerogel structural correlation radius, $\xi_0\approx20\text{--}80$\,nm depending on pressure \cite{VW}. This theory allows for a much smaller region of nonlinearity than a prior model, providing a better fit for our results \cite{sur25,dmit25}.

\section{Samples and methods}
The experimental setup replicated our previous work \cite{bet,dmit23,dmit25}, utilizing a nuclear demagnetization cryostat pre-cooled by a dilution refrigerator to achieve temperatures $\sim1$\,mK. Central to the measurements was a Stycast-1266 epoxy cell mounted on the cryostat's top stage, containing both a vibrating wire (VW) resonator and quartz tuning fork for thermometry. A superconducting solenoid outside the cell generated magnetic fields up to 31\,kOe with 0.7\% homogeneity over the $\pm$3\,mm sample region.

The nematic aerogel sample used in the experiments was made of mullite (Al$_2$O$_3\cdot$SiO$_2$), manufactured by Metallurg Engineering Ltd. The sample had a bulk density of approximately 3.1\,g/cm$^3$, a porosity of 95.2\%, and a final density of 150\,mg/cm$^3$. The aerogel strands had a diameter of $\leq$14\,nm, and the characteristic distance between strands was 60\,nm. Due to the anisotropy of the aerogel, the mean free path of $^3$He quasiparticles depended on the direction: it was 900\,nm parallel to the strands and 235\,nm perpendicular to the strands at the $T\rightarrow0$ limit \cite{dmit19}. The sample had a shape of cuboid with dimensions of approximately $2\times3$\,mm transverse to the strands and 2.6\,mm along the strands. The sample was glued to a 240\,$\mu$m NbTi wire using Stycast-1266 epoxy. The wire was bent into an arch shape with a total height of 10\,mm and leg separation of 4\,mm. The vibrating wire oscillated in the direction parallel to the strands, with a resonance frequency in vacuum at $T\sim1$\,K of $f_0=621$\,Hz and the full width at half-maximum (FWHM) of 0.3\,Hz.

To stabilize the polar and $\beta$ phases, the aerogel strands were pre-plated with $\approx3$ atomic layers of $^4$He \cite{dmit18}. For experiments studying the splitting of the A phase in nematic aerogel we used isotopically pure $^3$He, so the strands were instead covered by a natural thin layer of solid $^3$He.

In pure $^3$He, measurements were performed in magnetic fields $H=5\text{--}31$\,kOe and at pressure of 7.1\,bar. For the case of $^4$He coverage, measurements were conducted in magnetic fields $H=1\text{--}11$\,kOe at pressures of 7.1 and 19.4\,bar. During the experiments, an alternating current was applied to the wire, with amplitude $I_0$ varying from 0.85 to 2\,mA and from 0.4 to 4.4\,mA depending on $H$ for the cases of pure $^3$He and $^4$He coverage respectively. The current amplitude was set to ensure field-independent oscillation amplitude ($HI_0=const$). In liquid $^3$He the wire maximum velocity in the used temperature range was $\le0.2$\,mm/s not overheating $^3$He in and outside the aerogel sample. In the magnetic field produced by the main solenoid, the wire oscillated perpendicular to this field, generating an electromotive force (EMF). EMF signal was amplified by a step-up transformer 1:30 and measured using a lock-in amplifier. By slowly warming up the system and scanning the frequency, resonance curves were recorded and approximated by Lorentzian fits. These fits were used to construct the temperature dependence of the resonance parameters, allowing determination of the superfluid transition temperatures.

\begin{figure}[]
\includegraphics[width=\columnwidth]{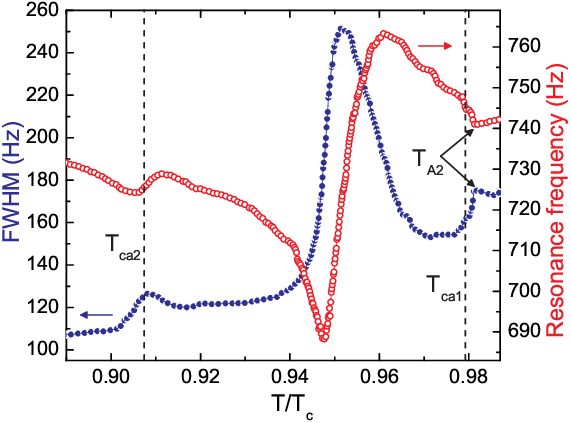}
\caption{Fig.\,\thefigure. (Color online)
Temperature dependence of FWHM (filled symbols, left axis) and the resonance frequency (open symbols, right axis) of the main mode of the aerogel VW without $^4$He layers at pressure of 15.4\,bar in magnetic field of 30.7\,kOe. Dashed lines indicate the superfluid transition temperatures in aerogel. Arrows indicate A$_1$-A$_2$ transition in bulk $^3$He ($T_{A2}$). The x-axis represents the temperature normalized to the superfluid transition temperature in bulk $^3$He, $T_c=2.083$\,mK.}
\label{fig1}
\end{figure}

\begin{figure}[]
\includegraphics[width=\columnwidth]{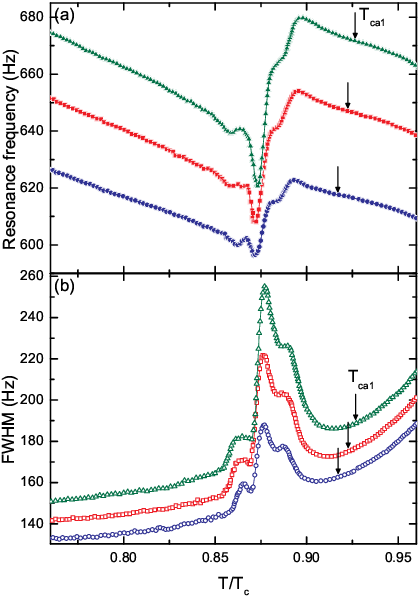}
\caption{Fig.\,\thefigure. (Color online)
Temperature dependence of the resonance frequency (a) and FWHM (b) of the main mode of the aerogel VW without $^4$He layers at pressure of 7.1\,bar in magnetic fields of 21.4\,kOe (circles), 27.2\,kOe (squares), and 30.7\,kOe (triangles). For clarity, the squares and triangles in the panel (a) are down-shifted by 60\,Hz and 80\,Hz respectively. Arrows indicate the upper critical temperature. $T_c=1.643$\,mK.}
\label{fig2}
\end{figure}

\begin{figure}[]
\includegraphics[width=\columnwidth]{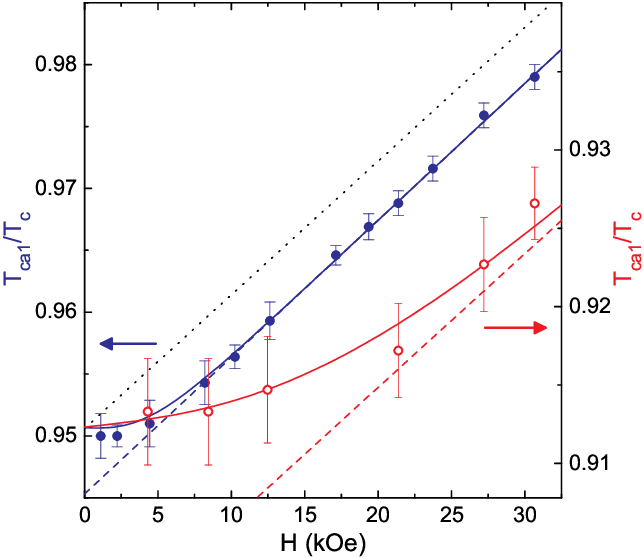}
\caption{Fig.\,\thefigure. (Color online)
The upper superfluid transition temperature in nematic aerogel without $^4$He coverage at pressures of 15.4\,bar (left scale, filled symbols, $T_{ca}\approx0.951T_c$) and 7.1\,bar (right scale, open symbols, $T_{ca}\approx0.912T_c$). The dotted line represents the bulk upper phase transition for both pressures scaled to their corresponding $T_{ca}/T_c$ (coincide due to the chosen axis scaling) \cite{isr84}. For both pressures, the solid lines are the best fit of our data with Eq.~\eqref{tca12}, where $\eta_1^0$ was set to the value expected for bulk A$_1$ phase \cite{isr84}. For 7.1\,bar $\alpha_0\approx0.047$\,kOe$^{-1}$ was fixed according to the theory as well, while for 15.4\,bar $\alpha_0$ was a fitting parameter resulting in $\alpha\approx0.215$\,kOe$^{-1}$ which is basically 6 times higher than its expected value $\alpha\approx0.039$\,kOe$^{-1}$. The dashed lines are asymptotes of the high-field data. Data for 15.4\,bar are from \cite{dmit25}.}
\label{fig3}
\end{figure}

\begin{figure}[]
\includegraphics[width=\columnwidth]{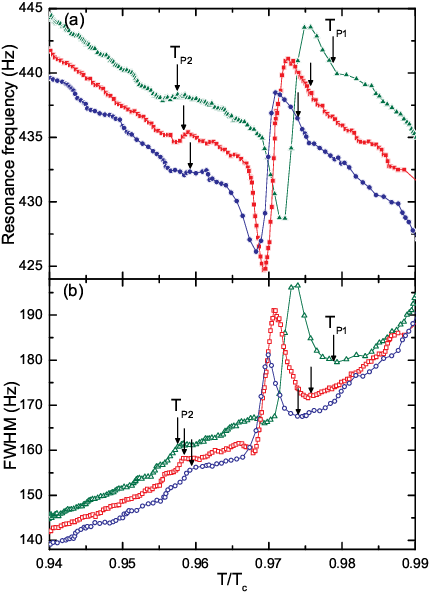}
\caption{Fig.\,\thefigure. (Color online)
Temperature dependence of the resonance frequency (a) and FWHM (b) of the main mode of the aerogel VW with $^4$He coverage at pressure of 7.1\,bar in magnetic fields of 6.7\,kOe (circles), 8.9\,kOe (squares), and 11.1\,kOe (triangles). For clarity, the squares and triangles in the panel (a) are down-shifted by 14\,Hz and 31\,Hz respectively, the triangles in the panel (b) are up-shifted by 5\,Hz. Arrows indicate the superfluid transition temperatures in aerogel.}
\label{fig4}
\end{figure}

\begin{figure}[]
\includegraphics[width=\columnwidth]{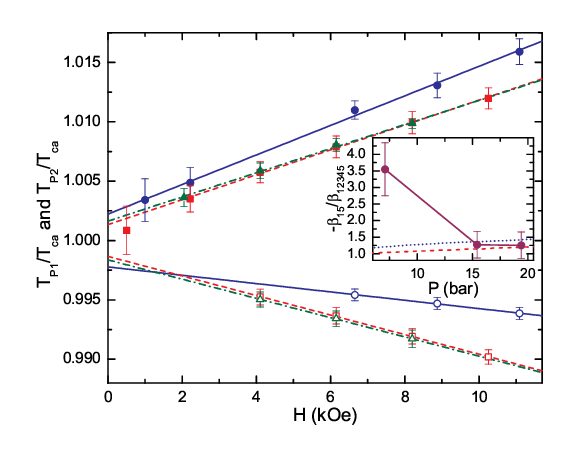}
\caption{Fig.\,\thefigure. (Color online)
The upper (filled symbols) and lower (open symbols) transition temperatures in nematic aerogel with $^4$He coverage at pressures of 7.1\,bar (circles, $T_{ca}\approx0.963T_c$), 15.4\,bar (squares, $T_{ca}\approx0.980T_c$), and 19.4\,bar (triangles, $T_{ca}\approx0.985T_c$) with their corresponding linear approximations. Here, $T_{ca}\equiv\left(T_{P1}\left|_{H=0}\right.+T_{P2}\left|_{H=0}\right.\right)/2$. The linear fits do not match at $H=0$ presumably due to a systematic error caused by finite temperature width of the superfluid transition. The inset shows the dependence of the ratio of slopes of the fit lines obtained from our experimental data defined as $-\beta_{15}/\beta_{12345}$ (see Eq.\,\eqref{beta}), the same ratio calculated using data from \cite{halp13} obtained in experiments with bulk $^3$He (dotted line) and with $^3$He in 98\% porosity silica aerogel (dashed line).}
\label{fig5}
\end{figure}

\begin{figure}[]
\includegraphics[width=\columnwidth]{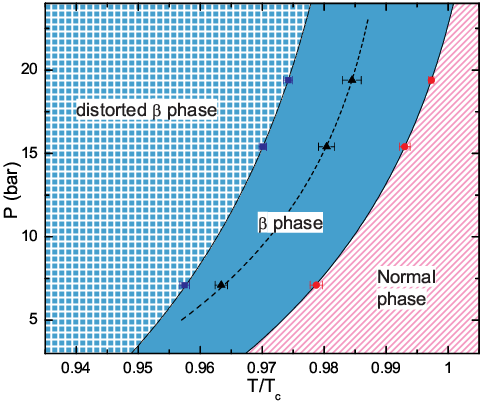}
\caption{Fig.\,\thefigure. (Color online)
The phase diagram of superfluid $^3$He in nematic aerogel under magnetic field of 11.1\,kOe. Circles and squares indicate the transition temperatures between the normal and $\beta$ phases, the $\beta$ and distorted $\beta$ phases of $^3$He respectively. Triangles represent $T_{ca}$ defined as in caption to Fig.~\ref{fig5}. Solid and dashed lines are guides to eye.}
\label{fig6}
\end{figure}

\section{Results and discussion}
\subsection{In pure $^3$He}
To determine the superfluid transition temperature, we measured the temperature dependence of the resonance parameters of the VW resonator in different magnetic fields. Fig.~\ref{fig1} shows the temperature dependence of the resonance frequency (open symbols) and FWHM (filled symbols) for the main mode of the VW resonance in the absence of $^4$He layers in a magnetic field of 30.7\,kOe obtained in our previous experiments at 15.4\,bar \cite{dmit25}. The superfluid transition in the aerogel occurs at $T=T_{ca1}$ into the A$_1$ phase, followed by a transition into the A$_2$ phase at lower temperatures. A peak-like behavior of FWHM around $T\approx0.95T_c$ is attributed to the interaction between the main mode of the VW and an additional (second) resonance mode, which corresponds to the oscillations of the superfluid and normal components of $^3$He within the aerogel and is analogous to the second sound \cite{dmit20sec,sur22}. The upper transition temperature $T_{ca1}$ was determined by deviations from the linear temperature dependence of the resonance parameters in the normal phase of $^3$He in aerogel, as described in previous studies \cite{dmit23}. At $T=T_{ca2}$ we observed a peak in the resonance width and a kink in the frequency. The height of this peak and its position changed when varying the magnetic field, so we associated this feature with the lower transition temperature $T_{ca2}$.

For a lower pressure of 7.1\,bar we conducted similar experiments. Fig.~\ref{fig2} shows the results in magnetic fields of 21.4\,kOe, 27.2\,kOe and 30.7\,kOe. In this case, the transition temperature into the A$_1$ phase was lower, around $T\approx0.91T_c$. However, the features previously associated with the transition to the A$_2$ phase were no longer distinguishable both for the frequency and FWHM data, in any field. Consequently, for this pressure, only the field dependence of $T_{ca1}$ was reliably determined using the same technique, but with noticeably higher error bars.

Fig.~\ref{fig3} presents the upper superfluid transition temperature in nematic aerogel without $^4$He coverage. The dotted line indicates the bulk dependency for each pressure, while solid lines represent fits to our data based on theoretical predictions of Baramidze and Kharadze \cite{bh}. According to theory, the suppression of the transition temperature can be attributed to scattering of $^3$He quasiparticles on the aerogel strands coated with paramagnetic solid $^3$He, acting as magnetic impurities. In our previous work at 15.4\,bar \cite{dmit25} we have found that $T_{ca1}$ exhibits $\approx6$ times smaller region of nonlinearity (around 5\,kOe) than expected, determined by $\alpha$. At higher fields it saturates and becomes linear, with the same slope as in the bulk phase with the same zero-field critical temperature. This much narrower region of nonlinearity contradicts the theory, but can be explained by the other effect of magnetic correlations proposed by Surovtsev \cite{sur25}. As for 7.1\,bar, we obtain $T_{ca1}$ with a rather worse accuracy and the saturation is not obvious in the used fields. The fit having correct $\alpha_0$ seems to be acceptable, but not sufficient enough to tell that the spin-polarized impurity model works here for sure.

\subsection{With $^4$He coating}
Similar measurements of the temperature dependence of the VW resonance parameters were also performed in the aerogel with $^4$He-coated strands. Fig.~\ref{fig4} shows the results of the resonance frequency (panel (a)) and FWHM (panel (b)) measurements at a pressure of 7.1\,bar in magnetic fields of 6.7\,kOe (circles), 8.9\,kOe (squares), and 11.1\,kOe (triangles). The peak-like behavior of the resonance width in this case is also explained by the interaction with the second oscillation mode \cite{dmit20sec}. Unlike in pure $^3$He, the temperature of the upper transition $T_{P1}$ was designated as the minimum in the FWHM versus temperature graph to be consistent with our original work with the $\beta$ phase \cite{bet}, because the results differed only systematically by the value of the error bar from those obtained using the FWHM versus resonance frequency graph (as in the previous section). The lower transition occurring at $T=T_{P2}$ is indicated by the kink in the frequency versus temperature dependence and by the step in the temperature dependence of FWHM. So for sufficient accuracy, $T_{P2}$ was determined by horizontally shifting temperature dependencies of FWHM for all pressures so that the step coincides, and from the value of the shifts we calculated the corresponding $T_{P2}$.

Measurements of the splitting with $^4$He coverage are shown in Fig.~\ref{fig5} for pressures of 7.1\,bar (circles), 15.4\,bar (squares), and 19.4\,bar (triangles). The linear fits reveal that the splitting becomes more symmetric as the pressure increases. At 7.1\,bar, the ratio of slopes $(dT_{P1}/dH)/(-dT_{P2}/dH)$ equals 3.5, significantly larger than the expected values for bulk $^3$He and $^3$He in silica aerogel. Also, the slope of the lower superfluid transition is smaller resulting in a rather bad overall accuracy, unlike at 15.4\,bar and 19.4\,bar, where this ratio is within the acceptable range.

Fig.~\ref{fig6} demonstrates the phase diagram of $^3$He in nematic aerogel under magnetic field of 11.1\,kOe. The transitions between the normal and $\beta$ phases, as well as between the $\beta$ and distorted $\beta$ phases, are indicated. For 15.4\,bar and 19.4\,bar the data were extrapolated to 11.1\,kOe.

\section{Conclusion}

In this work, we investigated the splitting of the superfluid transition temperature in $^3$He confined in nematic aerogel under varying pressures, with and without $^4$He coverage. Our experiments revealed a nonlinear dependence of the upper superfluid transition temperature $T_{ca1}$ on the magnetic field in the absence of $^4$He coverage. This nonlinearity spanned across a wider range of fields at lower pressure, while at higher pressure a clear transition to linear regime was observed. Notably, the observed region of nonlinearity at higher pressure is significantly narrower than predicted by existing theories based on magnetic scattering. We attribute this discrepancy to correlations in the magnetization of impurities, which affect the transition temperature into the A$_1$ phase of superfluid $^3$He, highlighting the importance of collective effects in the impurity system.

Since this effect does not depend on the anisotropy of the aerogel structure, we hypothesize that it should also be observed in isotropic samples. Future experiments using isotropic aerogel could provide a direct test of this hypothesis and further clarify the role of impurity correlations in the superfluid transition of $^3$He.

For the case of $^4$He coverage, the splitting of the transition temperature exhibited a more symmetric field dependence, with the ratio of slopes for the upper and lower transitions aligning more closely with theoretical predictions for bulk $^3$He.

The aerogel sample was made and provided by M.S.~Kutuzov from Metallurg Engineering Ltd. We are grateful to I.A.~Fomin and E.V.~Surovtsev for useful discussions and comments.

\section*{Funding}
This work was supported by ongoing institutional funding. No additional grants to carry out or direct this particular research were obtained.

\section*{Conflict of interest}
The authors of this work declare that they have no conflict of interest.

\end{document}